\documentclass[aps,superscriptaddress]{revtex4}

\usepackage[dvipdfm]{graphicx}
\usepackage{amsmath, amssymb, cases}

\usepackage{color}

\usepackage{ulem}

\begin{document}
\title{Sufficient conditions for wave instability in three-component reaction-diffusion systems}
\author{Shigefumi Hata}
\affiliation{Department of Physical Chemistry, Fritz Haber Institute of the Max Planck Society, Faradayweg 4-6, 14195 Berlin, Germany.}
\author{Hiroya Nakao}
\affiliation{Department of Mechanical and Environmental Informatics, Tokyo Institute of Technology, Ookayama 2-12-1, 152-8552 Tokyo, Japan}
\author{Alexander S. Mikhailov}
\affiliation{Department of Physical Chemistry, Fritz Haber Institute of the Max Planck Society, Faradayweg 4-6, 14195 Berlin, Germany.}

%%%%%%%%%%%%%%%%%%%%%%%%%%%%%%%%%%%%%%%%%%%%%%%%%%%%%%%%%%%%%%%%%%
\begin{abstract}
Sufficient conditions for the wave instability in general three-component reaction-diffusion systems are derived.
These conditions are expressed in terms of the Jacobian matrix of the uniform steady state of the system, and enable us to determine whether the wave instability can be observed as the mobility of one of the species is gradually increased.
It is found that the instability can also occur if one of the three species does not diffuse.
Our results provide a useful criterion for searching wave instabilities in reaction-diffusion systems of various origins.
\end{abstract}
%%%%%%%%%%%%%%%%%%%%%%%%%%%%%%%%%%%%%%%%%%%%%%%%%%%%%%%%%%%%%%%%%%

\maketitle

%%%%%%%%%%%%%%%%%%%%%%%%%%%%%%%%%%%%%%%%%%%%%%%%%%%%%%%%%%%%%%%%%%

%\section{Introduction}
The wave instability provides an important mechanism for pattern formation in nonequilibrium chemical systems.
When it takes place, a critical mode corresponding to a traveling wave with a certain wavenumber and oscillation frequency begins to grow,
destabilizing the uniform steady state.
Although being less known, the wave instability has already been considered in 1952 by A. Turing in his pioneering publication~\cite{Turing1952},
where the classical (i.e. static) Turing instability,
leading to the establishment of a periodic stationary pattern has also been introduced.
Therefore, it may be also appropriate to describe it as the oscillatory Turing bifurcation.
Moreover, it was already noticed by A. Turing~\cite{Turing1952} that at least three interacting species are needed for this instability to occur. 

Because of the spatial reflection symmetry, waves traveling in the left and right directions have the same growth rates
and both of them begin to spontaneously develop above the instability threshold.
Nonlinear interactions between such modes determine whether one of the modes gets suppressed,
so that a wave traveling in a certain direction is established,
or standing waves, representing superpositions of left and right traveling waves, are instead formed~\cite{Knobloch1990, Walgraef1996}.
The wave patterns resulting from such instability can also exhibit further instabilities, and wave turbulence may set on. 

In contrast to the classical Turing bifurcation,
which has been extensively discussed for both biological and chemical systems
\cite{Turing1952, Sick2006, Kondo1995, Nakamasu2009, Castets1990, Ouyang1991},
the wave bifurcation has so far attracted less attention.
It has been considered for special chemical models~\cite{Yang2002}
and its existence was suggested in the experiments with Belousov-Zhabotinsky microemulsions~\cite{Vanag2001}.
There are also publications in which this instability was discussed for special ecological models~\cite{Wang2007}.

Because at least three species are needed for the wave instability to occur,
the linear stability analysis is more complex in this case,
as compared with the classical Turing bifurcation in two-component activator-inhibitor systems.
The complexity of the stability analysis, which has been previously performed separately for individual chemical systems,
has probably also been responsible for the fact that the wave instability has not been broadly investigated for reaction-diffusion media. 

In this article, we present a general derivation of the sufficient conditions for the wave (or oscillatory Turing) bifurcation in arbitrary three-component reaction-diffusion models.
The final conditions are formulated in terms of the elements of the Jacobian matrix of the uniform steady state.
They tell whether the wave bifurcation is possible when the mobility of any chosen species is gradually increased,
while diffusion coefficients of other species are kept constant.
As we show, the instability is possible even if one of the three species is immobile.
Using our general expressions, the analysis of the wave bifurcation in specific reaction models will be largely simplified.

%%%%%%%%%%%%%%%%%%%%%%%%%%%%%%%%%%%%%%%%%%%%%%%%%%%%%%%%%%%%%%%%%%

%\section{Turing wave instabilities in three-component reaction-diffusion systems}
We consider reaction-diffusion systems with three chemical reactants $U, V$ and $W$.
Local densities of the reactants are denoted as $u = [U], v = [V]$ and $w = [W]$.
All reactants diffuse over the space and undergo local chemical reactions.
Generally, such systems are described by equations
\begin{align}
\begin{cases}
\displaystyle \dfrac{du}{dt} = \displaystyle f(u,v,w) +  D_{u} \nabla^{2}u,\\[8pt]
\displaystyle \dfrac{dv}{dt} = \displaystyle g(u,v,w) +  D_{v} \nabla^{2}v,\\[8pt]
\displaystyle \dfrac{dw}{dt} = \displaystyle h(u,v,w) +  D_{w} \nabla^{2}w,
\label{eq01}
\end{cases}
\end{align}
where functions $f, g$ and $h$ represent the local reactions.
Diffusion coefficients of the reactants are $D_{u}, D_{v}$ and $D_{w}$.
We assume that a uniform steady state $(u, v, w) = (\bar u, \bar v, \bar w)$ determined by
$f(\bar u, \bar v, \bar w) = g(\bar u, \bar v, \bar w) = h(\bar u, \bar v, \bar w) = 0$ exists
and that this state is stable in absence of diffusion.

We introduce small perturbations to the steady state as $(u, v, w) = (\bar u, \bar v, \bar w) + (\delta u, \delta v, \delta w)$.
Substituting this into Eqs.~(\ref{eq01}), the following linearized differential equations for the perturbations are obtained:
\begin{align}
\begin{cases}
\displaystyle \frac{d}{dt}\delta u = f_{u}\delta u + f_{v}\delta v + f_{w}\delta w +  D_{u} \nabla^{2} \delta u,\\[8pt]
\displaystyle \frac{d}{dt}\delta v = g_{u}\delta u + g_{v}\delta v + g_{w}\delta w +  D_{v} \nabla^{2} \delta v,\\[8pt]
\displaystyle \frac{d}{dt}\delta w = h_{u}\delta u + h_{v}\delta v + h_{w}\delta w +  D_{w} \nabla^{2} \delta w,
\label{eq02}
\end{cases}
\end{align}
where $f_u = \partial f / \partial u |_{(\bar u, \bar v, \bar w)}$,
$f_v = \partial f / \partial v |_{(\bar u, \bar v, \bar w)}$,
$f_w = \partial f / \partial w |_{(\bar u, \bar v, \bar w)}$,
... are partial derivatives at the steady state.
The following rescaled variables are introduced for convenience:
\begin{align}
\delta \tilde u = \delta u, \quad
\delta \tilde v = \sqrt{\left | \frac{ f_{v} }{g_{u}} \right |} \delta v, \quad
\delta \tilde w = \left | \frac{f_{v} }{h_{v}} \right |\delta w, \quad
\tilde t = \sqrt{\left | f_{v} g_{u} \right | }t.
\label{eq03}
\end{align}
We substitute these variables into Eqs.~(\ref{eq02}) to obtain a set of equations
\begin{align}
\begin{cases}
\displaystyle \frac{d}{d \tilde t}\delta \tilde u = m_{0}\delta \tilde u + \alpha \delta \tilde v + n \delta \tilde w +  D_{u} \mu \nabla^{2} \delta \tilde u,\\[8pt]
\displaystyle \frac{d}{d \tilde t}\delta \tilde v = \beta \delta \tilde u + p_{0}\delta \tilde v + q\delta \tilde w +  D_{v} \mu \nabla^{2} \delta \tilde v,\\[8pt]
\displaystyle \frac{d}{d \tilde t}\delta \tilde w = r \delta \tilde u +\gamma \delta \tilde v + s_{0}\delta \tilde w +  D_{w} \mu \nabla^{2} \delta \tilde w,
\label{eq04}
\end{cases}
\end{align}
where
\begin{align}
& m_{0} = \frac{ f_{u} }{\sqrt{\left | f_{v} g_{u} \right | }}, \qquad 
n = \frac{ f_{w} \left | h_{v} \right |}{\sqrt{\left | {f_{v}}^{3} g_{u} \right |}},\cr
& p_{0} = \frac{g_{v}}{\sqrt{ \left | f_{v} g_{u} \right | }},\qquad 
q = g_{w} \left | \frac{ h_{v}}{ f_{v} g_{u} } \right |,\cr
& r = \frac{h_{u}}{ \left | h_{v} \right |} \sqrt{ \left | \frac{ f_{v} }{g_{u}} \right | },\qquad 
s_{0} = \frac{ h_{w} }{\sqrt{ \left | f_{v} g_{u} \right | }}, \cr
& \mu = \frac{1}{\sqrt{\left | f_{v} g_{u} \right | }}.
\label{eq04}
\end{align}
The coefficients $\alpha, \beta, \gamma$ are determined by signs of $f_{v}, g_{u}, h_{v}$, $\alpha = \textrm{sign}(f_{v}), \beta = \textrm{sign}(g_{u}), \gamma = \textrm{sign}(h_{v})$.
The perturbations $(\delta \tilde u,\delta \tilde v,\delta \tilde w)$ are expanded over plane waves as
\begin{align}
\delta \tilde u = & \int d\vec k \tilde U^{\left( \vec k \right)} \exp \left [\lambda ^{\left( \vec k \right)} \tilde t -i\vec k \cdot \vec x \right ],\cr
\delta \tilde v = & \int d\vec k \tilde V^{\left( \vec k \right)} \exp \left [\lambda ^{\left( \vec k \right)} \tilde t -i\vec k \cdot \vec x \right ],\label{tmp}\\
\delta \tilde w = & \int d\vec k \tilde W^{\left( \vec k \right)} \exp \left [\lambda ^{\left( \vec k \right)} \tilde t -i\vec k \cdot \vec x \right ],\nonumber
\end{align}
where $\vec k = (k_{1}, k_{2}, \cdots)$ is the wave vector,
and $\lambda ^{\left( \vec k \right)}$ is a growth rate of the plane wave with wave vector $\vec k$.
Thus, we obtain the following equations for each wave vector $\vec k$:
\begin{equation}
\lambda^{(k)}
\begin{pmatrix}
\tilde U^{(k)}\\
\tilde V^{(k)}\\
\tilde W^{(k)}
\end{pmatrix}
=
\begin{pmatrix}
m_{0} - D_{u} \mu k^{2}& \alpha & n\\
\beta & p_{0} - D_{v} \mu k^{2} & q\\
r & \gamma & s_{0} - D_{w} \mu k^{2} 
\end{pmatrix}
\begin{pmatrix}
\tilde U^{(k)}\\
\tilde V^{(k)}\\
\tilde W^{(k)}
\end{pmatrix},
\label{eq06}
\end{equation}
where $k=\left |\vec k \right|$ is the wave number, the magnitude of the wave vector $\vec k$.
Because only the magnitude of the wave vector is important, here and below we drop the vector symbols.

The condition
\begin{equation}
\textrm{det}
\begin{pmatrix}
m_{0} - D_{u} \mu k^{2} - \lambda^{(k)}& \alpha & n\\
\beta & p_{0} - D_{v} \mu k^{2}- \lambda^{(k)} & q\\
r & \gamma & s_{0} - D_{w} \mu k^{2} - \lambda^{(k)}
\end{pmatrix}
=0
\label{eq_det}
\end{equation}
should be satisfied for Eq.~(\ref{eq06}) to have non-trivial solutions.
Thus, the linear growth rate $\lambda^{(k)}$ is determined by the characteristic equation
\begin{equation}
\lambda^{3} - (m +p +s) \lambda^{2} + (mp + ps + sm -nr -\gamma q -\alpha \beta)\lambda - \left ( mps -npr + \alpha qr - \gamma mq  + \beta \gamma n -\alpha \beta s \right ) = 0,
\label{eq07}
\end{equation}
where
\begin{align}
m = & m(k) = m_{0} - k^{2}\mu D_{u},\cr
p = & p(k) = p_{0} - k^{2}\mu D_{v},\label{eq08}\\
s = & s(k) = s_{0} - k^{2}\mu D_{w}.\nonumber
\end{align}

The growth of each plane-wave mode is determined by the real part of $\lambda^{(k)}$.
The uniform steady state is stable if $\textrm{Re}(\lambda^{(k)})$ is negative for all $k$.
The instability occurs if $\textrm{Re}(\lambda^{(k)})$ becomes positive for at least one wave number $k = k_\textrm{c}$.
Then the uniform steady state is destabilized, leading to spontaneous development of wave patterns with critical wave number $k_\textrm{c}$.
If the imaginary part $\textrm{Im}(\lambda^{(k_\textrm{c})})$ of the unstable mode is zero, the first critical mode represents a stationary plane wave and the Turing instability occurs.
On the other hand, if $\textrm{Im}(\lambda^{(k_\textrm{c})})\neq 0$, the critical mode is oscillatory in time and periodic in space,
so that the critical mode represents a traveling wave and the wave instability takes place.

%%%%%%%%%%%%%%%%%%%%%%%%%%%%%%%%%%%%%%%%%%%%%%%%%%%%%%%%%%%%%%%%%%
%\section{Critical condition for the instabilities}

The {\it complex conjugate root theorem} tells that a characteristic equation with real coefficients
\begin{align}
\lambda^{3} + a \lambda^{2} + b \lambda + c = 0
\label{tmp2}
\end{align}
has either three real roots or one real root and a pair of complex conjugate roots.
In the latter case, the three roots can be written as
\begin{equation}
\lambda_{1, 2} = \psi \pm i \omega, \quad
\lambda_{3} = \phi,
\label{eq09}
\end{equation}
so that the coefficients are represented as
\begin{equation}
a = - (2\psi + \phi),\quad
b = \psi^{2} + \omega^{2} + 2 \psi \phi, \quad
c = - (\psi^{2} + \omega^{2})\phi.
\label{eq10}
\end{equation}
Combining these three equations, we obtain
\begin{align}
c - ab = & - (\psi^{2} + \omega^{2})\phi + (2\psi + \phi) (\psi^{2} + \omega^{2} + 2 \psi \phi)\cr
= & 2\psi \left [ (\psi + \phi)^{2} + \omega^{2} \right ].
\label{eq11}
\end{align}

At the threshold of the wave instability, we have $\psi = 0$, so that $c-ab=0$.
At the threshold of the Turing instability, we would have $\phi = 0$, and therefore $c = 0$.
Note that the Turing instability is also possible when the characteristic equation has three real roots and one of them becomes positive.
It can be easily checked that, also in this case, the instability threshold corresponds to $c=0$.

Thus, the wave instability first takes place when, for one wavenumber $k = k_{\textrm c}$, the equation
\begin{align}
I_{\textrm{wav}}= & ( mps -npr + \alpha qr - \gamma mq  + \beta \gamma n -\alpha \beta s ) + (m +p +s) (mp + ps + sm -nr -\gamma q -\alpha \beta) =0
\label{eq12}
\end{align}
becomes satisfied, where $m, p$ and $s$ are given by equations~(\ref{eq08}) with $k = k_{\textrm c}$.

The Turing instability first takes place when, for one wavenumber $k = k_{\textrm c}$, the equation
\begin{align}
I_{\textrm{st}}= & mps -npr + \alpha qr - \gamma mq  + \beta \gamma n -\alpha \beta s =0
\label{eq13}
\end{align}
becomes satisfied, where the coefficients $m,p$ and $s$ are again given by equations~(\ref{eq08}) with wavenumber $k_{\textrm c}$.

It is convenient to introduce the three-dimensional $m$-$p$-$s$ space in order to represent 
these conditions graphically.
As illustrated in Figures 1 and 2, each of the conditions~(\ref{eq12}) and~(\ref{eq13}) defines a boundary surface
and Eqs.~(\ref{eq08}) determine a straight line $\Gamma$ which is parameterized by the wave number $k$.
If the line $\Gamma$ touches the boundary surface $I_{\textrm{wav}}(m,p,s)= 0$ or $I_{\textrm{st}}(m,p,s)= 0$,
then the conditions~(\ref{eq12}) or~(\ref{eq13}) are satisfied and the corresponding instability takes place.

%%%% fig.1 %%
\begin{figure}[t]
\begin{center}
\includegraphics[width=0.85\hsize]{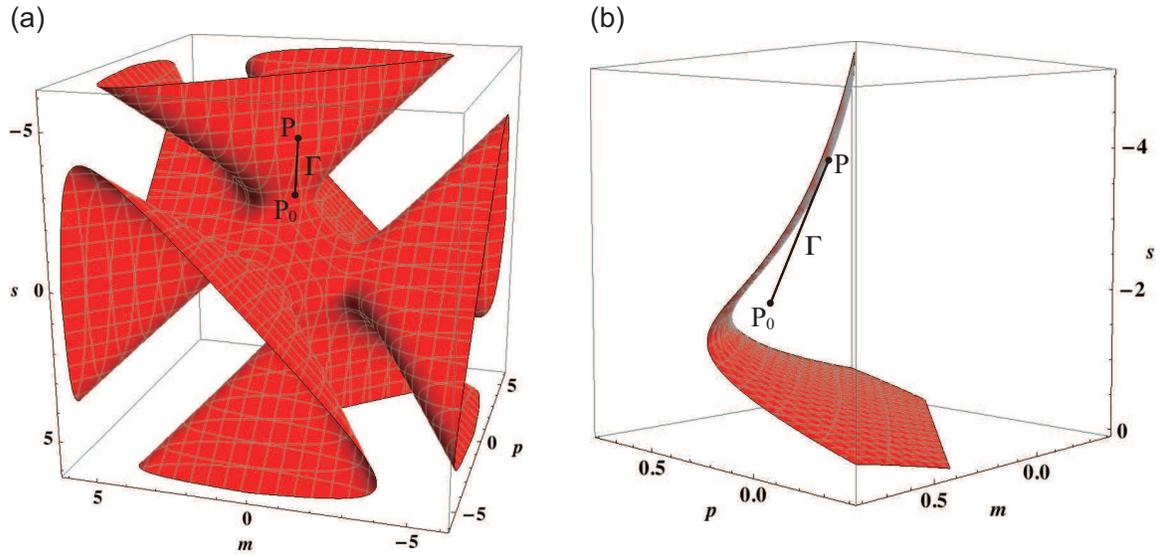}
\caption{
Boundary surface for the wave instability $I_{\textrm{wav}}(m,p,s)=0$.
The line $\Gamma$ touches the surface at a point $P$.
Panel (b) is a closeup of (a).
Parameters are fixed at $n=q=r=\alpha=\beta=\gamma=1$, $m_{0}=0.4, p_{0}=0.4$ and $s_{0}=-1.8$.
Diffusion constants are $D_{u}=1, D_{v}=1$ and $D_{w}= 14.521$.
}
\label{fig01}
\end{center}
\end{figure}
%%

%%%% fig.2 %%
\begin{figure}[t]
\begin{center}
\includegraphics[width=0.45\hsize]{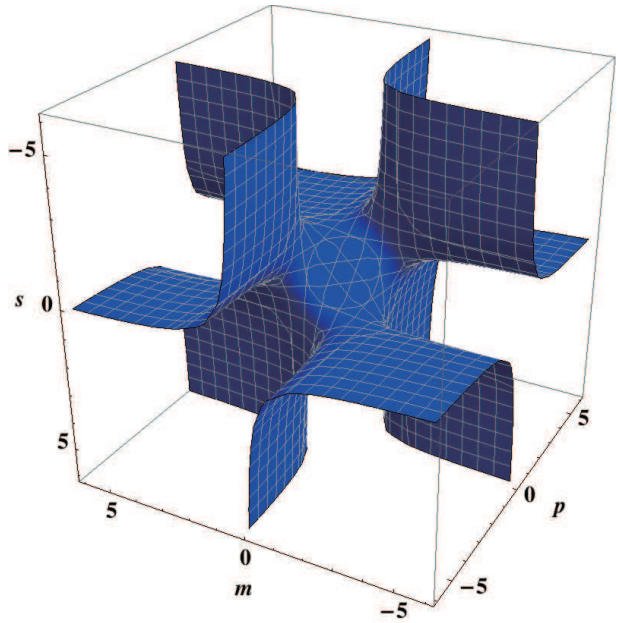}
\caption{
Boundary surface for the Turing instability $I_{\textrm{st}}(m,p,s)=0$.
Parameters are fixed at $n=q=r=\alpha=\beta=\gamma=1$.
}
\label{fig02}
\end{center}
\end{figure}
%%

%%%%%%%%%%%%%%%%%%%%%%%%%%%%%%%%%%%%%%%%%%%%%%%%%%%%%%%%%%%%%%%%%%
%\section{Necessary condition for the wave instability}

The uniform steady state $(\bar u, \bar v, \bar w)$ should be stable if the diffusion is absent.
In terms of of the Jacobian matrix at the steady state
\begin{align}
J =
\begin{pmatrix}
f_{u} & f_{v} & f_{w}\\
g_{u} & g_{v} & g_{w}\\
h_{u} & h_{v} & h_{w}
\end{pmatrix},
\label{Jac01}
\end{align}
this implies that
\begin{align}
\textrm{det}(J)<0 \quad \textrm{and} \quad \textrm{tr}(J)<0.
\label{Jac02}
\end{align}
Thus, the initial point $P_{0}$ with coordinates $(m_{0},p_{0},s_{0})$ on the line~(\ref{eq08}) should lie inside the stable region in $m$-$p$-$s$ space.

%%%%%%%%%%%%%%%%%%%%%%%%%%%%%%%%%%%%%%%%%%%%%%%%%%%%%%%%%%%%%%%%%%
%\section{Sufficient condition for the wave instability}
The wave instability takes place if the first critical mode is oscillatory.
At the threshold of the wave instability, 
the line $\Gamma$ defined by Eq.~(\ref{eq08}) should touch the boundary surface $I_\textrm{wav}=0$ without having intersections with the surface $I_\textrm{st}=0$.

Suppose that we want to check whether the wave instability can occur when the diffusion constant $D_{w}$ is varied.
Let us denote $A = D_{v} / D_{u}$ and $B = p_0 - m_0 D_{v} / D_{u}$ and consider a plane $p = A m + B$ parallel to the $s$-axis.
The line $\Gamma$ is always lying on this plane irrespective of $D_w$, because the conditions $p_{0} = A m_{0} + B$ and $D_{v} / D_{u} = A$ are satisfied.
A coordinate $x$ is introduced on the plane in such a way that we have $x=0$ when $m+p=0$ holds.
The slope of the line $\Gamma$ in the $x$-$s$ space is $D_{u}D_{w} / \sqrt{{D_{u}}^{2}+{D_{v}}^{2}}$ and is nonnegative.
The initial point $P_{0}$ of the line $\Gamma$ is at $(x_{0},s_{0})$ on the plane, where $x_{0} = (m_{0}+p_{0})/ (A+1)$.
Because we assume that the necessary conditions~(\ref{Jac02}) are satisfied,
the point $(x_{0},s_{0})$ is in the stable region of the $x$-$s$ space.
Two surfaces $I_\textrm{wav}=0$ and $I_\textrm{st}=0$ intersect the plane along the boundary curves $\hat I_{\textrm{wav}}=0$ and $\hat I_{\textrm{st}}=0$,
\begin{align}
\hat I_{\textrm{wav}}(x,s;A, B) = I_{\textrm{osc}}\left (x-\frac{B}{A+1},\ A x + \frac{B}{A+1},\ s \right ) =0,
\label{eq14}\\
\hat I_{\textrm{st}}(x,s;A, B) = I_{\textrm{st}}\left (x-\frac{B}{A+1},\ A x + \frac{B}{A+1},\ s \right ) =0.
\label{eq15}
\end{align}
Below we examine the dependences of the boundary curves $\hat I_{\textrm{wav}}=0$ and $\hat I_{\textrm{st}}=0$ on the parameters
$m_0, p_0, s_0, n,q,r, \alpha, \beta$ and $\gamma$.
This allows us to obtain the parameter conditions under which the wave instability can occur.

\begin{figure}[t]
\begin{center}
\includegraphics[width=0.95\hsize]{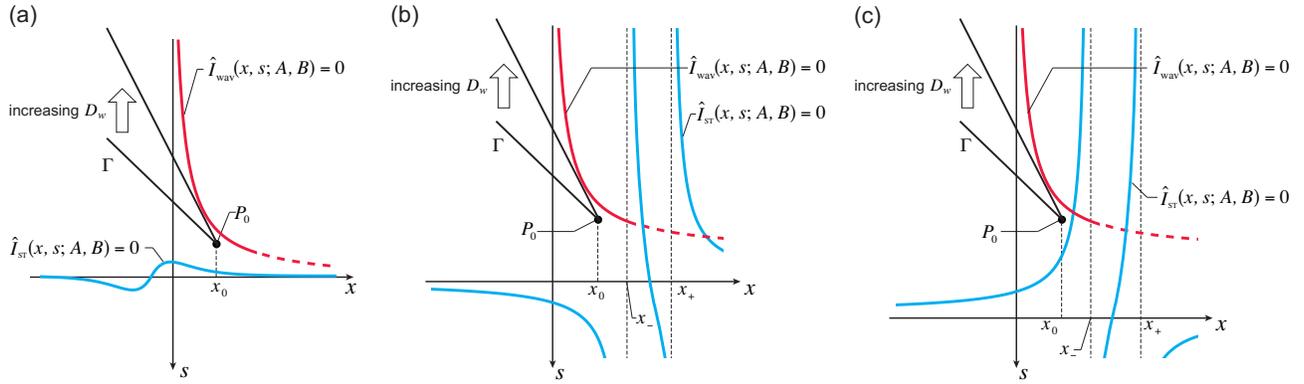}
\caption{
Boundary curves $\hat I_{\textrm{wav}} = 0$ (red) and $\hat I_{\textrm{st}} = 0$ (blue)
when (a) the condition~(\ref{eq21}) and (b,c) the condition~(\ref{eq22}) are satisfied.
All species are mobile.
Positions of the line $\Gamma$ defined by~(\ref{eq08}) are shown for two different values of $D_{w}$.
The boundary curve $\hat I_{\textrm{st}} = 0$ can behave as shown in (b) or (c), depending on the model parameters.
}
\label{fig03}
\end{center}
\end{figure}

\begin{figure}[t]
\begin{center}
\includegraphics[width=0.65\hsize]{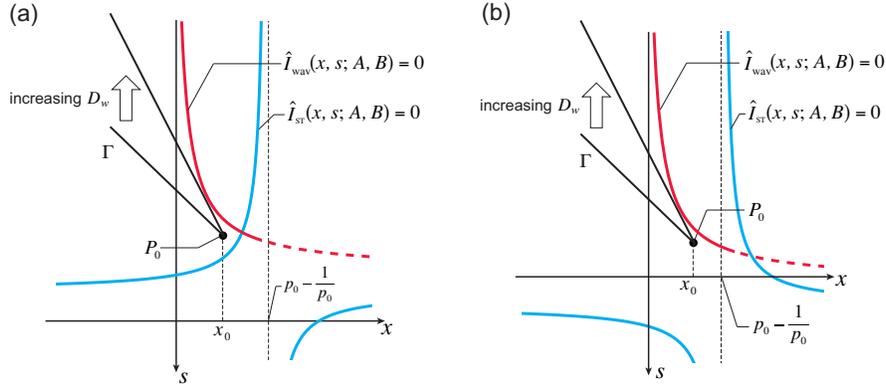}
\caption{
Boundary curves $\hat I_{\textrm{wav}} = 0$ (red) and $\hat I_{\textrm{st}} = 0$ (blue).
The reactant $V$ does not diffuse ($D_{v}=0$).
The line $\Gamma$ defined by~(\ref{eq08}) is shown as black lines for two different values of $D_{w}$.
The boundary curve $\hat I_{\textrm{st}} = 0$ can behave as shown in (a) or (b) depending on the parameters $n, q, r, \alpha, \beta$ and $\gamma$.
}
\label{fig04}
\end{center}
\end{figure}

%%%%%%%%%%%%%%%%%%%%%%%%%%%%%%%%%%
%\subsection{Boundary curve $\hat I_{\textrm{wav}}=0$}
Let us examine the shape of the boundary curve $\hat I_{\textrm{wav}}=0$ at large values of $|s|$.
If $|s|\gg 1$,
we can neglect $\mathcal{O}(s^{0})$ terms and obtain
\begin{align}
\hat I_{\textrm{wav}}(x,s;A, B) \approx s\left [ (1+A)^{2}x^{2} + s(1+A)x - nr -\gamma q \right ].
\label{eq16}
\end{align}
Then, the boundary curve $\hat I_{\textrm{wav}}=0$ is given by the equation
\begin{align}
(1+A)^{2}x^{2} + s(1+A)x - nr - \gamma q= 0
\label{eq165}
\end{align}
and we have
\begin{align}
x  = & \frac{1}{2} \left [ -\frac{s}{A+1} \pm \sqrt{\left(\frac{s}{A+1}\right)^{2}+4\frac{nr+\gamma q}{(A+1)^{2}}}\right ]\cr
\simeq  & \frac{s}{2(A+1)} \left [ -1 \pm \left ( 1+2\frac{nr+\gamma q}{s^{2}} \right ) \right ].
\label{eq17}
\end{align}
As follows from Eq.~(\ref{eq16}), the boundary curve approaches $x=0$ in the limit of large $|s|$ as
\begin{align}
(A+1)xs^2=0.
\label{eq175}
\end{align}
Then, the plus sign should be chosen in Eq.~(\ref{eq17}).
Thus, the asymptotic boundary curve in the limit $s \gg1$ is the hyperbola
\begin{align}
xs = \frac{nr + \gamma q}{A+1}.
\label{eq176}
\end{align}
If the coefficients of this hyperbola is negative, i.e. $nr+\gamma q <0$, the boundary curve lies in the region of $x>0$ and $s<0$.
In such cases, given that $x_0>0$, the line $\Gamma$ always touches the boundary curve under increasing the diffusion constant $D_w$,
as shown in Figs.~\ref{fig03} and~\ref{fig04}.
Therefore, we require $nr+ \gamma q<0$ and $x_0>0$ as a part of the sufficient conditions.
Using definitions of the model parameters~(\ref{eq04}), these requirements can be written in terms of the Jacobian matrix as
\begin{align}
f_{w}h_{u}+ g_{w}h_{v}<0,
\label{eq18}\\
f_u + g_v >0.
\label{eq19}
\end{align}
%%

%%%%%%%%%%%%%%%%%%%%%%%%%%%%%%%%%%
%\subsection{Boundary curve $\hat I_{\textrm{st}}=0$}
The boundary curve for the Turing instability $\hat I_{\textrm{st}}(x,s;A, B) = 0$ is given by
\begin{align}
s(x) = -\dfrac{(A+1)^{2}(Anr + \gamma q)x - (A+1)^{2}(\beta \gamma n + \alpha qr) + (A+1)B(nr-\gamma q)}{A(A+1)^{2}x^{2}-(A^{2}-1)Bx -\alpha \beta (A+1)^{2}-B^{2}},
\label{eq20}
\end{align}
which is a single-valued function of $x$.

%%%% fig.2 %%
%%

%% no diverge %%%%%%%%%%%%%%%%%%%%
%\subsubsection{Three diffusible reactants}
Let us assume that all three reactants diffuse over the space, $D_u\neq 0, D_v\neq 0$ and $D_w\neq 0$, which leads to $A\neq 0$.
If the denominator on the right hand side of Eq.~(\ref{eq20}) is not zero for all values of $x$,
then $s(x)$ is a continuous function.
Figure~\ref{fig03} (a) illustrates the qualitative shape of the boundary curves $\hat I_{\textrm{wav}}=0$ and $\hat I_{\textrm{st}}=0$ in such a situation.
In this case, the line $\Gamma$ touches the boundary curve $\hat I_{\textrm{wav}} = 0$
without intersecting $\hat I_{\textrm{st}} = 0$ as $D_w$ is increased.
As a consequence, the wave instability takes place.
The conditions are given by
\begin{align}
(A^{2}-1)^{2}B^{2}+4A(A+1)^{2}\left [ \alpha \beta (A+1)^{2}+B^{2} \right ] <0
\label{eq205}
\end{align}
which implies
\begin{align}
f_u g_v - f_v g_u > \dfrac{(f_u D_v + g_v D_u)^2}{4D_u D_v}.
\label{eq21}
\end{align}

If the denominator on the right hand side of Eq.~(\ref{eq20}) vanishes at $x_-$ and $x_+$,
the boundary curve $\hat I_{\textrm{st}} = 0$ diverges in two ways depending on the values of $n, q, r, \alpha, \beta$ and $\gamma$ (Fig.~\ref{fig03} (b) and (c)).
In both cases, if $x_0 \leq x_-$,  the line $\Gamma$ touches $\hat I_{\textrm{wav}} = 0$ first.
Thus, if the condition
\begin{align}
x_0 \leq x_- = \dfrac{(A^2 -1)B - (A+1)^2 \sqrt{B^2 +4\alpha \beta A}}{2A (A+1)^2}
\label{eq215}
\end{align}
which is equivalent to
\begin{align}
f_u g_v - f_v g_u > 0
\label{eq22}
\end{align}
and
\begin{align}
f_{u}D_{v}+g_{v}D_{u}\leq 0
\label{eq225}
\end{align}
is satisfied, the wave instability takes place as $D_w$ is increased.

%% 2 diffuse %%%%%%%%%%%%%%%%%%%%%
%\subsubsection{Two diffusible reactants}
Next, we assume that the reactant $V$ does not diffuse, that is $D_{v}=0$, leading to $A=0$.
In this case, the boundary curve $\hat I_{\textrm{st}} = 0$ is given by
\begin{align}
s(x)|_{A=0} = -\dfrac{q x + \kappa (n-qr) + B(nr-q)}{Bx + 1-B^{2}},
\label{eq22}
\end{align}
which diverges at $x = B-1/B = p_{0}-1/p_{0}$ [Figs.~\ref{fig04} (a) and (b)].
In this case, if $x_{0}\leq p_{0}-1/p_{0}$, the line $\Gamma$ touches $\hat I_{\textrm{wav}} = 0$
without intersecting $\hat I_{\textrm{st}} = 0$ when $D_{w}$ is increased.
Thus, the instability condition for a system with two diffusible reactants is given by
\begin{align}
x_{0}\leq p_{0}- \frac{1}{p_{0}},
\label{eq225}
\end{align}
which implies
\begin{align}
g_{v}(f_{u}g_{v} - f_v g_{u} ) \leq 0.
\label{eq23}
\end{align}
On the other hand, if the reactant $U$ does not diffuse, $D_{u}=0$, one can choose a plane $m = A' p + B'$ where $A' = D_{u}/ D_{v}$ and $B' = m_0 - p_0 D_{u} / D_{v}$ and derive the condition
\begin{align}
f_{u}(f_{u}g_{v} - f_v g_{u} ) \leq 0.\label{eq24}
\end{align}
%%

%%%% fig.3 %%

Thus, sufficient conditions for the wave instability under increasing diffusion constant $D_w$ of the reactant $W$ has been derived
for the four cases depending on the diffusion constants.
The first two conditions~(\ref{eq18}) and~(\ref{eq19}) are common to all cases.
The last condition depends of the diffusion constants, i.e. we have
\begin{numcases}{}
f_u g_v - f_v g_u > \dfrac{(f_{u}D_{v}+g_{v}D_{u})^{2}}{4D_{u}D_{v}} \quad (D_{u}\neq 0, D_{v}\neq 0),
\label{condi01}\\[9pt]
f_u g_v - f_v g_u >0 \quad \textrm{and} \quad f_{u}D_{v}+g_{v}D_{u}\leq 0 \quad (D_{u}\neq 0, D_{v}\neq 0),
\label{condi02}\\[9pt]
f_{u}(f_{u}g_{v}-f_{v}g_{u}) < 0 \quad (D_{u}=0),
\label{condi03}\\[9pt]
g_{v}(f_{u}g_{v}-f_{v}g_{u}) < 0 \quad (D_{v}=0).
\label{condi04}
\end{numcases}
These different requirements can however be expressed in a single equation;
\begin{align}
\textrm{det}
\begin{pmatrix}
f_u+D_{u} \Lambda & f_{v}\\
g_{u}& g_{v}+ D_{v} \Lambda
\end{pmatrix}
\neq 0 \ \textrm{for any} \ \Lambda<0.
\label{cond_5}
\end{align}
with at least two non-vanishing diffusion constants.

The wave instability may take place also under the variation of other two diffusion constants $D_u$ or $D_v$.
The corresponding sufficient conditions for each case can be obtained by permutating three variables $u,v$ and $w$.

In summary, the wave instability occurs under increasing diffusion constants $D_u, D_v$ or $D_w$, if the conditions
\begin{align}
g_{v}+h_{w}>0\quad \textrm{and} \quad g_{u}f_{v}+h_{u}f_{w}<0 \quad \textrm{and} \quad
\textrm{det}
\begin{pmatrix}
g_v+D_{v} \Lambda & g_{w}\\
h_{v}& h_{w}+ D_{w} \Lambda
\end{pmatrix}
\neq 0 \ \textrm{for any} \ \Lambda<0,
\label{condi07}
\end{align}
\begin{align}
h_{w}+f_{u}>0\quad \textrm{and} \quad h_{v}g_{w}+f_{v}g_{u}<0 \quad \textrm{and} \quad
\textrm{det}
\begin{pmatrix}
h_w+D_{w} \Lambda & h_{u}\\
f_{w}& f_{u}+ D_{u} \Lambda
\end{pmatrix}
\neq 0 \ \textrm{for any} \ \Lambda<0,
\label{condi08}
\end{align}
or
\begin{align}
f_{u}+g_{v}>0\quad \textrm{and} \quad f_{w}h_{u}+g_{w}h_{v}<0 \quad \textrm{and} \quad
\textrm{det}
\begin{pmatrix}
f_u+D_{u} \Lambda & f_{v}\\
g_{u}& g_{v}+ D_{v} \Lambda
\end{pmatrix}
\neq 0 \ \textrm{for any} \ \Lambda<0.
\label{condi06}
\end{align}
are satisfied.

%%%%%%%%%%%%%%%%%%%%%%%%%%%%%%%%%%
%\section{Summary}
%% summary
We have derived sufficient conditions for the wave instability in general three-component reaction-diffusion systems.
The conditions are formulated in terms of the Jacobian matrix elements at a steady state and of the diffusion constants
They do not depend on model details.
Once these conditions are satisfied, the wave instability occurs as we increase the diffusion mobility of one of the reacting species.

Our general results are applicable for systems of various origins,
including biological, chemical, physical and ecological systems.
Our analysis has revealed that the wave instability may occur even if one of three reactants is immobile.
This result can be important in a variety of applications involving both diffusible and non-diffusible reactants.

%%%%%%%%%%%%%%%%%%%%%%%%%%%%%%%%%%
%\section{Acknowledges}
Authors acknowledge the financial support through the DFG SFB 910 program ``Control of Self-Organizing Nonlinear Systems'' in Germany,
through the Fellowship for Research Abroad, KAKENHI and the FIRST Aihara Project (JSPS),
and the CREST Kokubu Project (JST) in Japan.
%% bib %%%%%%%%%%%%%%%%%%%%%%%%%%%%%%%%

%%%%%%%%%%%%%%%%%%%%%%%%%%%%%%%%%%%%%%%%%%%%%%%%%%%%%%%%%%%%%%%%%%%%

\end{document}